# First tests of a 800 kJ HTS SMES

P. Tixador, M. Deleglise, A. Badel, K. Berger, B. Bellin, J.C. Vallier, A. Allais, C.E. Bruzek

*Abstract*— SMES using high critical temperature superconductors are interesting for high power pulsed sources. Operation at temperatures above 20 K makes cryogenics easier, enhances stability and improves operation as pulsed power source. In the context of a DGA (Délégation Générale pour l'Armement) project, we have designed and constructed a 800 kJ SMES. The coil is wound with Nexans conductors made of Bi-2212 PIT tapes soldered in parallel. The coil consists in 26 superposed simple pancakes wound and bonded on sliced copper plates coated with epoxy. The rated current is 315 A for an energy of 814 kJ. The external diameter of the coil is 814 mm and its height 222 mm. The cooling at 20 K is only performed by conduction from cryocoolers to make cryogenics very friendly and invisible for the SMES users. The cooling down has been successfully carried out and the thermal system works as designed. After a brief description of the SMES design and construction, some tests will be presented. From a current of 244 A, the SMES delivered 425 kJ to a resistance with a maximum power of 175 kW.

*Index Terms*—Cryogenics, SMES, magnet, PIT tape.

## I. INTRODUCTION

A shorted-circuited superconducting (SC) magnet stores energy in the magnetic field created by a circulating current: the current remains constant due to the absence of resistance of superconductors [1]. The stored energy is transferred by opening the short-circuit on the load. It is the dual of a capacitor. This device is known under its acronym, SMES (Superconducting Magnetic Energy Storage).

The specific energy is limited by mechanical considerations (Viriel Theorem [2]) to values around some tens kJ/kg, which is lower than batteries but higher than most of power capacitors. The mechanical aspect of a SMES is consequently of the highest importance and the magnet conductor must be designed to support high stresses and deformations without SC property degradation.

The specific power has no theoretical limit and can be extremely high (100 MW/kg) which makes SMES a promising candidate for pulsed power sources either in the military or in the civil fields.

Manuscript received August 29, 2007. This work is supported by the DGA.
P. Tixador, M. Deleglise, A. Badel, K. Berger and J.C. Vallier are with the CNRS-Institut Néel/G2Elab, Grenoble, F-38 042 France (phone: 33-476-88-7949; e-mail: pascal.tixador@grenoble.cnrs.fr).
A. Allais is with Nexans France, F-92587 Clichy Cedex, France (e-mail: arnaud.allais@nexans.com).
B. Bellin was with CNRS and is now with, JST Transformers, F-69371 Lyon Cedex 08, France (e-mail: Boris.Bellin@jst-transformers.com).
C.E. Bruzek was with Nexans, he is now with LNE, F-78197 Trappes Cedex, France (e-mail: Christian-Eric.Bruzek@lne.fr).

SMES generally use NbTi magnets operating at 4 K. Some High Temperature Superconductor (HTS) SMESs have been developed [3, 4]. HTS, by increasing the operating temperature, makes the refrigeration system simpler and lowers exploitation and investment costs. The magnet stability with respect to external disturbances is well improved. The power (voltage x current) can be enhanced as higher temperature margins allow to increase the electrical isolation thickness, and therefore the operating voltage. AC loss removal is less problematic when the operating temperature increases.

## II. SMES DESIGN

A SMES of some hundreds kJ operating at 20 K with an integrated cryogenics is an intermediate and essential step to develop the basic technologies for manufacturing HTS SMES [5]. The purpose is to qualify several technological solutions on a representative level and to acquire an indispensable operation experience feedback. The two key elements are a HTS conductor adapted to SMES and an invisible cryogenics for the user, i.e. without cryogenic fluid handling.

### A. Superconducting conductor

The conductor basic element is a Bismuth PIT tape with the 2212 stoichiometry ($Bi_2Sr_2CaCu_2O$ – Bi-2212) produced by Nexans [6]. This tape imposes an operation around 20 K. Fig. 1 shows its critical characteristic at 20 K for two field directions: longitudinal and transversal. The tape is very sensitive to transverse magnetic field: to carry the same current, the number of tape will have to be higher in the areas with strong transverse field component than in those with longitudinal component.

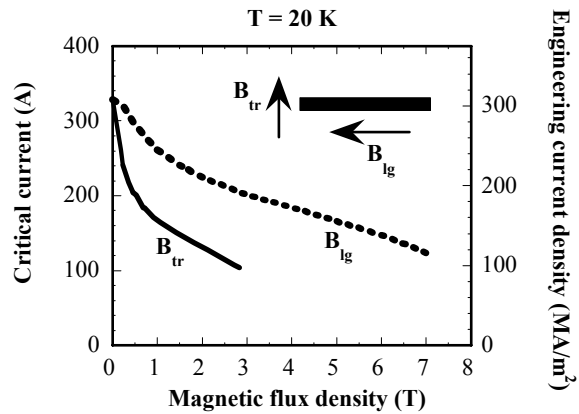

Fig. 1. Tape reference critical characteristics at 20 K under longitudinal ($B_{lg}$) and transverse ($B_{tr}$) magnetic flux densities.

The Bi-2212 filaments (33 % in volume) are embedded in a pure Silver matrix with an external sheath in AgMg in order to improve the mechanical properties. The allowable stress for pure silver is very low after the heat treatments necessary to form the SC phase. The critical stress is some tens of MPa and does not suit for SMES. The critical stress is the value for which the critical current decreased in a reversible way by 5%. Beyond this limit, degradation becomes fast and irreversible in general. Even with the AgMg sheath, the mechanical performances remain rather low (100 MPa) and a mechanical reinforcement is required in the areas with strong mechanical stresses.

The current for a pulsed SMES must be very large to reach high powers. For this first realization a current of approximately 300 A was chosen. Since the current capacity of a tape is on the order of hundred Amperes (Fig. 1), several tapes are put in parallel. We chose to solder them on top of each others. This configuration favorably allows current sharing between the tapes in case of a local defect on a tape. In addition, this structure makes it possible to adapt the tape number to the local flux densities (amplitude and direction) to optimize the use of superconductor. A stainless steel tape (Fig. 2) improves the allowable stress of the conductor. The critical stress reachs 170 MPa. Kapton® glued around the conductor with or without the reinforcement tape guarantees the electrical isolation (Fig. 2).

But with this conductor configuration, the tapes are electromagnetically coupled and behave like a large single tape. The AC losses are consequently important under longitudinal field since these are proportional to the conductor thickness.

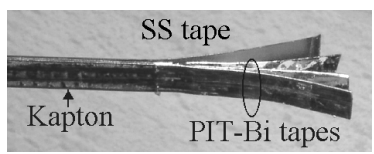

Fig. 2. Reinforced conductor for the SMES (4 mm width).

### B. Superconducting magnet

The SC magnet must be designed to minimize the SC material amount for a given magnetic energy, to ensure proper cooling and support of the electromagnetic forces. The magnet must also fulfill the specified electromagnetic signature. For this first realization, the leak field was not specified. Under these conditions, a solenoid is well adapted: the amount of superconductor is much less when compared to a torus. The realization and the support of the electromagnetic forces in a solenoid are also simpler than in a torus. The minimization of the SC volume leads to a rather flat solenoid even when considering the unfavorable effects of the transverse magnetic component, which is important in a flat solenoid. Pancake winding is appropriate to a flat solenoid. In addition pancakes offer an excellent mechanical holding and a very effective cooling since the exchange surfaces are then naturally very large. Within a pancake, it is also very easy to change the conductor by just soldering the two conductors one on top the other over a sufficient distance (higher than 50 mm in our case). Adapted wedges on both sides of the connection allow the continuous support of the conductor. The resistance of such an internal connection is about ten nΩ.

In a pancake, the absence of layer jump is favorable for PIT tapes, that are very sensitive to transverse deformations. On the other hand, connections between pancakes remain difficult. In particular, the inner connections are subjected to an important magnetic flux density that increases their resistivity (copper magnetoresistivity). Under 6 T, the resistance of our inner connections is only 0.15 μΩ, thanks to PIT tapes soldered between the two pancakes, on the copper connection. The technique of double pancakes without inner connection does not suit PIT tapes since it only accept very little transverse deformations.

The main characteristics of the magnet are shown in table 1. Fig. 3 indicates the conductor distribution. The reinforced conductor is in the center where the stress is maximum. The four tape conductor is used in extremity areas with strong transverse component of the magnetic field (Fig. 4). The three tape conductor is used elsewhere.

TABLE 1 Parameters of the 800 kJ Bi-2212 PIT SMES.

| Quantity | Value |
|---|---|
| Stored energy | 814 kJ |
| Internal / External coil diameter - Height | 300 mm / 814 mm – 222 mm |
| Rated current | 315 A |
| Operating temperature | 20 K |
| Number of pancakes | 26 |
| Max magnetic flux density (long/trans) | 5.2 T / 2.5 T |
| Max circumferential stress / axial stress | 80 MPa / 24 MPa |

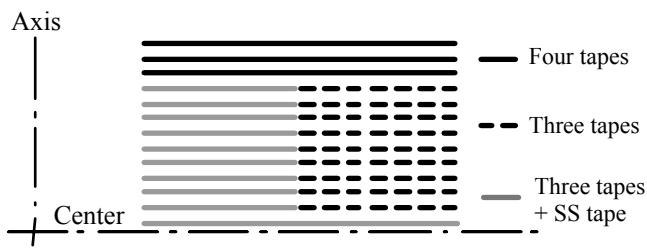

Fig. 3. Conductor distribution within half of the coil.

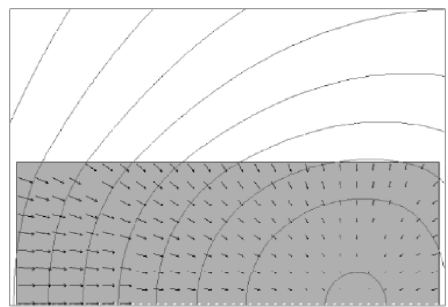

Fig. 4. Magnetic field distribution and forces (arrows, independent turns) within half of the coil.

At the rated energy, the conductor operates at 80 % of its critical current on the two load lines (longitudinal and transverse). This small margin is due to the present high cost of the PIT tapes. It leads to resistive losses in the magnet since the resistive transition index ("n" value) is about 15. A superior limit is given assuming that the total length operates at 80 % of its critical current: 13 W.

Fig. 5 show a stack of several pancakes on the under flange.

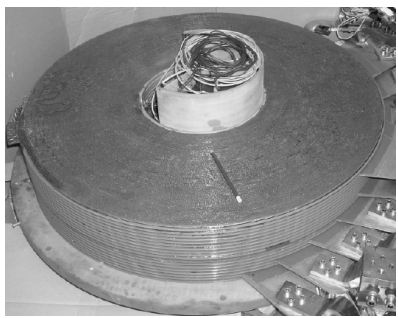

Fig. 5. The pancake magnet under assembling.

The maximum electromagnetic stress is 80 MPa. Taking into account the winding tensions and the thermal differential dilatations, the maximum stress is about 100 MPa. The turns in our magnet are not totally independent since they are glued on the same copper support, but they are not directly bonded together. Anyway, even if the electromagnetic stress distribution across a pancake is very different considering that turns act independently or dependently [7], the maximum value is the same (Fig. 6).

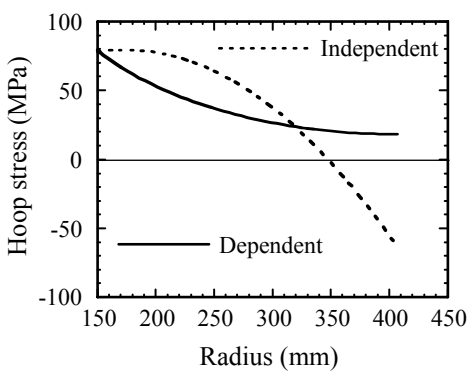

Fig. 6. Hoop stresses along the two central pancakes.

The current leads use Bi-2212 tubes from Nexans between the magnet and an intermediate point thermalized by a dedicated cryocooler, which also cools the thermal shield around the magnet. The upper part of the current leads uses brass. These HTS current leads clearly reduce the heat load on the extremity pancakes. They bring about 70 mW, which is still a lot for a conduction cooling. The current leads lower end is therefore thermalized to limit the temperature rise of the extremity connections.

*C. Cooling*

A conduction cooling using cryocoolers makes the cryogenics very friendly for the user, who does not have to handle any cryogenic fluid. This was one of the important specifications of the project. The thermal design must be particularly carefully studied to reduce the temperature gradients between the cryocooler and the magnet parts, especially the inner pancake connections that dissipate 20 mW with a conservative resistance value of 0.2 µΩ. Even if the use of HTS material increases the possible temperature gradients, those remain low, a few Kelvins maximum. The cryogenic design was specified with a maximum temperature difference of 3 K between the cryocooler and the hottest point, which is the inner connection the farthest of the cryocooler (AL 330 from Cryomech [8]). All the thermal design has been made by numerical simulations (software FLUX® [9]) associated with many measurements of thermal resistances [10] (pressed contacts, joinings, …).

To properly cool the pancakes, the conductor is bonded on a copper plate covered with epoxy resin for the electrical isolation. The bond is obtained after complete winding, by curing the epoxy film adhesive between the conductor and the copper plate. The thermal contact between conductor and Cu is good (measured conductance of 110 W/m$^2$/K at 20 K).

The Cu plate is cut out to reduce the eddy current losses during fast discharges of the magnet (Fig. 7). The length of inner cuttings is a compromise between the temperature gradient in steady state operation and the eddy current losses. Each plate has a wing connected by a flexible Cu part to the thermal distribution sector where the cryocooler is screwed (Fig. 7). The Cu connecting parts isolate partly the pancakes from the vibrations of the Gifford MacMahon cryocooler and allow differential dilatations. To benefit from an excellent thermal conductivity at 20 K, we used CuC1 quality copper.

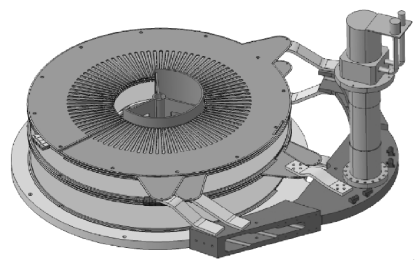

Fig. 7. Pancake Cu plates and their thermal connection to the Cu thermal distribution sector with the cryocooler.

A thermal shield (Cu and Al alloy) surrounds the magnet to lower the radiation losses. It also intercepts the losses of the mechanical supports (composite rods) of the magnet to the external flange. The second cryocooler cools this shield as well as the intermediate point of the HTC current leads. Superisolation is largely used to reduce the radiation losses. Fig. 8 shows the interior of the SMES with the magnet and the upper flange.

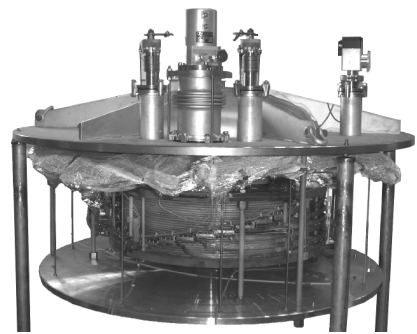

Fig. 8. Inner SMES picture with the 26 pancake magnet.

## III. SMES TESTS

The coil midpoint is linked to the earth to reduce the voltage on the pancakes. The quench detection is based on the imbalance between the high and low parts of the magnet. A potentiometer balances the two parts when the magnet is superconducting. The magnet protection is triggered when the imbalanced voltage oversteps the threshold values in amplitude and duration. Circuits using mutual inductances complete this quench detection but these show much less sensitivities in case of a quench.. Magnet protection consists in discharging the current in a resistor by opening the power supply connection (when charging) and the magnet short-circuit (when in storage mode). For safety reasons, the discharge resistor is always connected to the magnet

A first test of the SMES was carried out with 10 pancakes [5]. It validated the cryogenic design and the 10 pancakes were tested at their operating temperature. Nevertheless the mechanical stresses and the magnetic flux density were lower than in the final magnet with 26 pancakes. Several pancakes carried 300 A [5]. These tests had revealed two pancakes with a reduced critical current (150 A). They were rewound.

The second test campaign concerned the complete magnet. Approximately one week is necessary to completely cool the SMES. It takes practically 3.5 days to really stabilize the temperatures (20 K - 13 K) after a 3.5 day quick temperature decrease (300 K - 20 K). Without current, the minimal temperatures reached on the two cryocoolers are 11.7 K (magnet) and 15.8 K (shield and current leads) corresponding to cold powers of 3 W and 20 W respectively. These experimental values are in good agreement with the design. The 3 W value is not very accurate (accuracies of the temperature sensor and power-temperature characteristic of the cryocooler). The 20 W value mainly represents the two current lead heat load. The pancake temperatures, measured on their wings, are rather homogeneous and are about 13 K.

A low sinusoidal current at variable frequency first supplied the magnet. The losses were proportional to the frequency in the investigated frequency range (10 mHz – 20 Hz) at low currents (0.12 A). The maximum voltage of the power supply limits the current at "high" frequencies since the magnet inductance is large. This result confirms that the AC losses are mainly hysteresis losses and that the eddy current losses in the cryostat are low.

The 26 pancake magnet carried 200 A but during this test, the next to the extremity pancake quenched and was locally damaged since the detection delay was too long. This is in agreement with the hot spot temperature calculation. Fig. 9 shows the current and the pancake resistance evolution. The resistance development is slow and it makes the quick quench detection difficult. The resistance reached less than 5 % of the 300 K value after 30 s. This low value shows that the quench was very localized. After this test this pancake does not carry in steady state more than 50 A.
The protection system was improved to lower a lot the detection delay. We put a temporary current lead to shunt the damaged pancake. This removable current lead does not use HTS and is thermalized on the He vessel. This He vessel is under the thermal distribution sector and was foreseen during the design in case of insufficient cold power or too large thermal gradients. It is not used in normal operation. The tests at high currents are then carried out with 24 coils instead of 26. The self inductance is reduced by about 15 % and the rated energy is now about 700 kJ instead of 814 kJ.

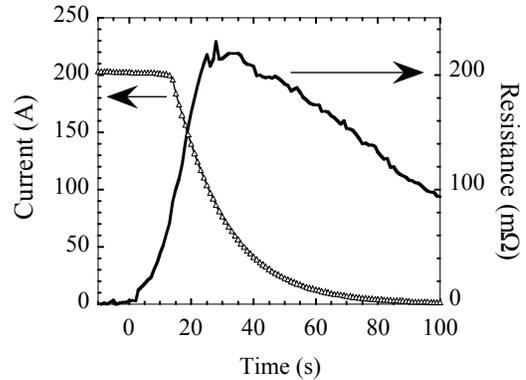

Fig. 9. Current and resistance evolutions during the quench of pancake 25.

Fig. 10 shows the discharge of the 24 coil SMES in a resistance starting from a current of 244 A. The discharge was triggered by the quench detector during the magnet energizing. The power reached 175 kW and the energy dissipated in the resistance was 425 kJ. The losses in the magnet are low since the initial magnetic energy was of the same order of magnitude. It is in agreement with calculations of the AC losses in the conductor and the eddy current losses in the cryostat.

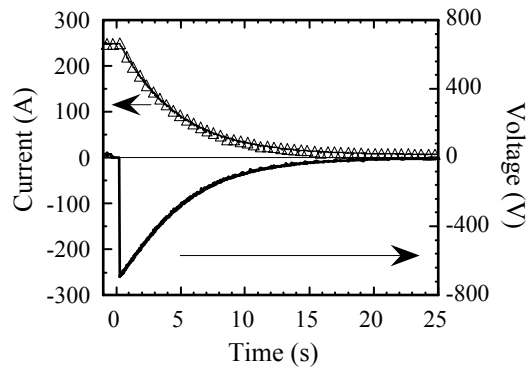

Fig. 10. Current and voltage during a resistive discharge.

Fig. 11 shows some temperature evolutions during this discharge. The initial temperatures are low since the He vessel cools not only the temporary current lead, but also the magnet. The temperature rise is limited (2 K) and a current of 240 A may be injected again in the magnet some minutes after the discharge.

Fig. 12 shows the discharge of the total magnet in a power capacitor. The association of capacitor and SMES may be interesting for high power sources. A power diode is inserted between the SMES and the capacitor to avoid oscillations. The voltage, therefore the current, were voluntarily limited. This quick (approximately 150 ms) discharge also shows low losses: the capacitor voltage follows in particular well the sinusoidal theoretical curve without any losses (Fig. 12). The capacitor voltage reduction after its charge is certainly due to the power diode, during its blocking.

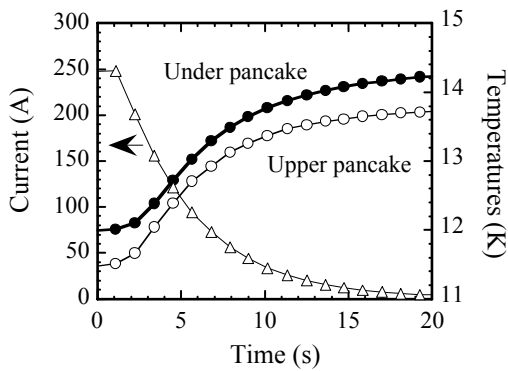

Fig. 11. Temperature evolution during a discharge.

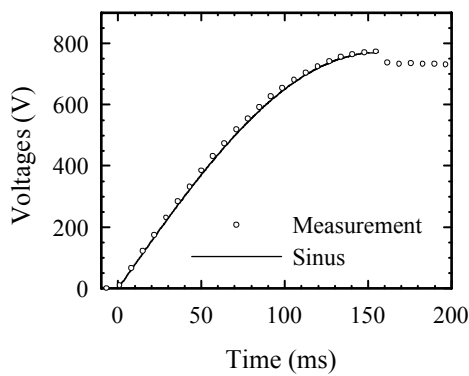

Fig. 12. Capacitor voltage during the SMES discharge.

## IV. CONCLUSION

The HTS SMES appears as a very interesting solution for pulsed power sources. If mass energy density is limited by mechanical considerations to satisfying values, the mass power density is definitely more interesting. HTSs bring many opportunities for these pulsed SMES. We have designed and constructed a 800 kJ SMES with Bi-2212 Nexans PIT conductor. The SMES is only cooled by solid conduction for the most invisible cryogenics possible. The thermal system operates globally in a satisfactory way and in agreement with the calculations. Nevertheless some temperature sensors show abnormal values, which could explain that some pancakes operate at higher temperatures than the designed value of 20 K. The tests carried on the magnet were conclusive even if the rated energy still has not been reached. For the current, 80 % of the rated value was recorded. Fast discharges show low losses in the magnet and its close environment. The rises in temperature in this case are very limited (2 K). The SMES will be dismounted for inspection and analysis. The coated conductors are very promising for pulsed SMES. Their use is in our perspectives.

ACKNOWLEDGMENT

The authors warmly thank A. Boulbes, G. Barthelemy, G. Andre, P. Chantib and D. Grand for their major and essential technical contribution. They also thank O. Exchaw, J.L. Bret and G. Simiand for the electronic protection system.